# Dynamic Spectrum Sharing Based on the Rentable NFT Standard ERC4907

Litao Ye
*College of Electronics and Information Engineering*
*Shenzhen University*
Shenzhen, China
yelitao210708856@163.com

Bin Chen*
*College of Electronics and Information Engineering*
*Shenzhen University*
Shenzhen, China
bchen@szu.edu.cn

Shrivastava Shivanshu
*Department of Electrical and Electronics Engineering*
*Rajiv Gandhi Institute of Petroleum Technology*
Amethi-229304, India
shvnshushrivastava88@gmail.com

Chen Sun
*Wireless Network Research Department*
*Research and Development Center, Sony (China) Limited*
Beijing, China
chen.sun@sony.com

Shuo Wang
*Wireless Network Research Department*
*Research and Development Center, Sony (China) Limited*
Beijing, China
shuo.wang@sony.com

Siming Feng
*College of Electronics and Information Engineering*
*Shenzhen University*
Shenzhen, China
1214118618@qq.com

Shengli Zhang
*College of Electronics and Information Engineering*
*Shenzhen University*
Shenzhen, China
zsl@szu.edu.cn

*Abstract*—Centralized Dynamic Spectrum Sharing (DSS) faces challenges like data security, high management costs, and limited scalability. To address these issues, a blockchain-based DSS scheme has been proposed in this paper. First, we utilize the ERC4907 standard to mint Non-Fungible Spectrum Tokens (NFSTs) that serve as unique identifiers for spectrum resources and facilitate renting. Next, we develop a smart contract for NFST auctions, ensuring secure spectrum transactions through the auction process. Lastly, we create a Web3 spectrum auction platform where users can access idle spectrum data and participate in auctions for NFST leases corresponding to the available spectrum. Experimental results demonstrate that our NFST, designed according to the ERC4907 standard, effectively meets users' secure and efficient DSS requirements, making it a feasible solution.

## I. INTRODUCTION

With the rapid advancements in new-generation network technologies, like 6G mobile communication networks and the Internet of Things (IoT), the demand for the already scarce spectrum has increased exponentially [1]. Most wireless networks usually adopt a static spectrum allocation scheme, where a large part of radio spectrum is allocated to a fixed licensee over a long term. However, static allocation often creates unused spectrum resources, known as spectrum holes, due to variations in spectrum utilization across different times, frequencies, and spaces [2]. Therefore, more flexible and dynamic spectrum allocation schemes are needed to optimize spectrum utilization.

Cognitive Radio (CR) technology [3] aims to solve the problems of spectrum scarcity and low spectrum utilization. CR allows Dynamic Spectrum Sharing (DSS) for Secondary Users (SUs). DSS enables the SUs to sense and use the unused spectrum authorized to the Primary User (PU), thereby improving spectrum utilization [4]. However, traditional DSS acquires idle spectrum in a centralized way, making it vulnerable to security attacks. Attackers tamper with spectrum data according to their interests, significantly threatening the data and asset security of both sides of the spectrum transaction [5]. In addition, when PUs lease unused idle spectrum to SUs, SUs pay a high-security deposit raising the leasing threshold [6]. Moreover, the complexity of the transaction contract, transaction costs, control of the lease duration, and the return of the spectrum usage right after expiration also face management difficulties, and these factors make PUs less motivated to share their idle spectrum [7]. Therefore, maintaining security, trustworthiness, and transparency in the DSS process has become crucial nowadays.

Applying blockchain technology to DSS offers new possibilities to solve the above problems. For instance, in [8], a blockchain architecture for DSS was proposed, presenting the system architecture of the Spectrum Blockchain, construction of the Spectrum Trusted Ledger, management of Spectrum Smart Contracts, and spectrum transaction settlement methods. However, the specific design approach for secure spectrum resource transactions through smart contracts was not detailed. In [6], a Multi-Ops Spectrum Sharing (MOSS) smart contract was designed to facilitate DSS, enhancing security and reliability compared to traditional methods. Nevertheless, an explanation for ensuring the uniqueness of the spectrum following a spectrum auction was not provided by the authors. Spectral Token [9], designed based on the Non-Fungible Token (NFT) standard ERC721 [10], is a solution. Spectral Token ensures the uniqueness of the spectrum and effectively reduces the probability of spectrum interference occurring. However, the authors did not consider that during the spectrum leasing period, the ownership of the Spectral Token has been transferred to the lessee, which may lead to accidents that are not under the control of the spectrum owner. In addition, upon expiration of the lease, the authors need to explain how Spectral Token ownership can be returned safely and efficiently.

In this paper, we address the above problems by creating spectrum tokens using the NFT standard ERC4907 [11]. This approach guarantees precise monitoring and documentation of spectrum usage and leasing activities, thereby ensuring accurate delineation of ownership and usage rights. Furthermore, our methodology incorporates an automatic reset mechanism for spectrum usage rights to their initial values after each lease cycle through the utilization of smart contracts. This not only enhances security but also streamlines the leasing process. Additionally, we introduce a spectrum sharing platform designed to facilitate the execution of smart contracts.

## II. Related Background

Smart Contract is an automated contract built on blockchain technology [12]. It is stored and executed on blockchain network nodes. Each node verifies the execution result through a consensus mechanism. Smart contracts can be utilized for implementing ERC721 token contracts to facilitate the creation and management of NFTs. NFTs possess unique identifiers and characteristics that prove asset ownership and uniqueness. However, when it comes to leasing NFTs minted under the ERC721 standard, certain limitations exist. To enable NFT leasing, complex smart contracts must be meticulously designed, resulting in significant gas consumption. Furthermore, security issues may arise during the leasing process. Hence, there is a pressing need to develop a simplified and secure protocol that enables efficient and safe NFT leasing.

ERC4907 is an extension of the ERC721 standard that introduces a dual-role concept comprising the "owner" and "user" roles. It aims to streamline the management of NFT leases and enhance lease security. Through ERC4907, NFT owners can securely transfer the right to use their NFT to a specific user and set the term of use without altering the NFT owner's address. Upon the completion of the lease term, there is no need for the NFT owner to initiate an on-chain transaction and consume Gas to modify the NFT user address. Instead, the NFT user address is automatically reset to its initial value, thereby minimizing the costs associated with the lease.

These features of smart contracts and the ERC4907 standard are suitable for DSS. Leveraging smart contracts enables the standardization of the entire spectrum transaction process, eliminating the need for trusted intermediaries and enhancing efficiency while reducing costs. Furthermore, the ERC4907 standard is employed to segregate spectrum owner and user. During the lease period, the SU assumes the role of spectrum user. After the lease expires, the user address is automatically reset to the initial value. The PU consistently retains ownership. This separation simplifies spectrum leasing and enhances security, deterring potential malicious activities. SUs are not required to make substantial security deposits to establish trustworthiness, significantly reducing the barrier to entry for spectrum leasing while incentivizing PUs to share their idle spectrum resources.

## III. System Model and Spectrum Leasing Process

### A. System Model

Blockchain technology offers decentralized DSS, making it secure and reliable without any trusted third party involvement. We apply the auction of NFSTs and lease periods minted based on ERC4907 standard to DSS. The overall system model is illustrated in Fig. 1, comprising key components such as blockchain, the Spectrum Management Authority (SMA), PUs, bidders, the minting module, and the auction module. Further elaboration on the system model is provided below.

a) Blockchain: Blockchain securely records, verifies, and stores spectrum data, providing a trusted execution environment for smart contracts.

b) SMA: Deploying smart contracts, uploading spectrum data to the blockchain, minting NFSTs for authorized PUs, and maintaining the spectrum auction platform.

c) PUs: Spectrum leasers, which own the NFSTs minted by the SMA.

d) Bidders: The bidders can be SUs or other PUs with a demand for spectrum usage. The PUs model is denoted as $PUs = \{PuBuyer_1, PuBuyer_2, ... , PuBuyer_i\}$. The SUs model is represented as $SUs = \{SU_1, SU_2, ... , SU_j\}$. Bidders participate in the auction by calling the bidding contract on the auction platform. The auction winner obtains the right to use the spectrum during the lease period.

e) Minting Module: MSA mints NFSTs for PUs based on the ERC4907 standard and spectrum-related parameters.

f) Auction Module: Users with spectrum demand participate in the auction for the NFST lease term set by the PU based on the NFST ID to complete the on-chain auction for the NFST.

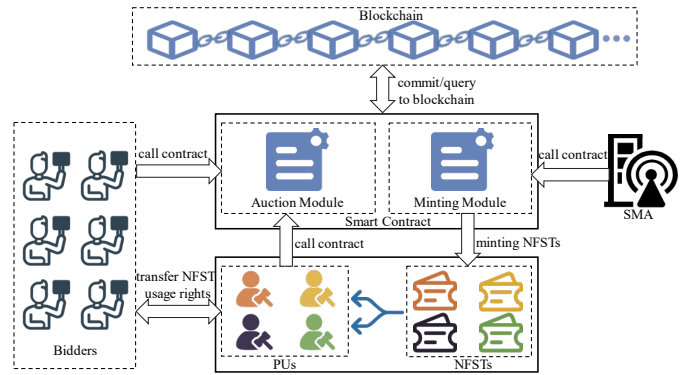

Fig. 1. System model.

### B. DSS Process

The SMA mints NFSTs for spectrum owners, adhering to the ERC4907 standard. These NFSTs are minted based on spectrum information such as spectrum geo-location, frequency bandwidth, and spectrum owner address. As illustrated in Fig. 2, the DSS process is followed when users require a spectrum. The flow of this process is outlined as follows:

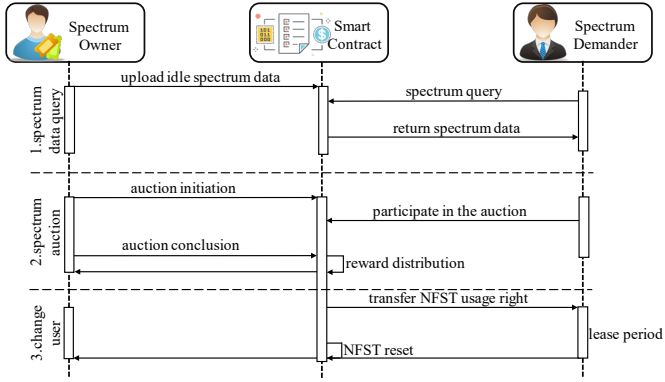

Fig. 2. Spectrum sharing process.

The spectrum sharing process consists of three key stages: spectrum data query, spectrum bidding, and spectrum use rights. Initially, the spectrum owner uploads available spectrum data to the blockchain. Subsequently, spectrum demanders access idle spectrum data through smart contracts. The spectrum owner initiates the bidding process, allowing spectrum demanders to participate in auctions by invoking the relevant contract with known spectrum data.

Following the auction's conclusion, the spectrum owner finalizes it through the contract, and the auction's proceeds are distributed as per the contract design. The contract automatically updates the address of the highest bidder, designating them as the spectrum user. During the lease period, the spectrum user can utilize the spectrum resource as per regulations. Upon lease expiration, the contract resets the spectrum user address to the initial address, completing a round of spectrum sharing.

## IV. NFST BASED ON ERC4907 STANDARD

### A. Minting of NFSTs

To address the challenges posed by collision and spectrum interference among users in centralized DSS systems, we employ the ERC4907 standard for minting NFSTs within the DSS system model discussed in Section III. The design of smart contracts is facilitated through the Remix integrated development environment (IDE) [13], which is subsequently deployed and executed on the Ethereum [14] network. A pseudo-code representation of the minting contract is depicted in Fig. 3, where the SMA uses input parameters, including the spectrum owner's address, frequency bandwidth, and spectrum geo-location, to generate NFSTs with unique identifiers.

```
mint_NFST // deployed and called by SMA
  input: owner, startFreq, endFreq, geoLocation
  if(caller ≠ SMA) then
      return False
  owner ← spectrum license holder address
  minAlloc ← minimum allocatable size of a frequency band
  freq ← startFreq
  while freq < endFreq
      mint a new NFST(owner: owner, startFreq: freq, endFreq: freq + minAlloc,
          geoLocation: geoLocation, issuer: signature of caller)
      freq ← ferq + minAlloc
  return True
```

Fig. 3. Minting NFSTs contract.

An instance of the NFST minting outcome is showcased in Fig. 4, with an ID value of "1" and the corresponding owner address identified as 0xdD870fA1b7C4700F2BD7f44238821C26f7392148.

Fig. 4. NFST minting results.

### B. NFST User Settings

At the auction's conclusion, the spectrum owner calls the auction-end contract. The smart contract automatically takes the winner's address and the expiration time as input parameters, following the preset code logic, and configures the leasing information of the NFST. At this time, the status of this spectrum is updated to "*Occupied*." Fig. 5 shows the pseudo-code of NFST user settings.

```
set_user // called by spectrum owner
  input: tokenId, highestBidder, lease duration
  if(caller ≠ spectrum owner) then
      return False
  user ← highestBidder
  // update spectrum status
  NFSTInfo[tokenId].status ← "Occupied"
  expires ← currentTimestamp + lease duration
  // set the user of NFST
  setUser(tokenId: tokenId, user: user, expires: expires, issuer: signature of caller)
  return True
```

Fig. 5. NFST user settings contract.

After the auction ends, the established procedures implement the NFST user settings. The smart contract ensures secure and controlled access to the leased spectrum resources using NFSTs by setting the winner's address and expiration time.

## V. SPECTRUM AUCTION PROCESS AND RESULTS

This section explains the pseudo-code design for the NFST lease period auction contract, bidder calls on the Web3 auction platform, and the bidding results.

### A. Spectrum Auction Process

Spectrum demanders access a Web3 Auction Platform to retrieve idle spectrum data and its corresponding NFST IDs. They then participate in the auction by entering the NFST ID and their bid amount for the desired spectrum. We employ the English Auction, allowing participants to increase their bids gradually. Fig. 6 illustrates the pseudo-code of the NFST auction contract.

```
start_auction // called by spectrum owner
  input: tokenId, auctionDuration, lease duration, beneficiary, startingPrice
  if(caller ≠ spectrum owner or userOf(tokenId) ≠ null) then
      return False
  endTime ← currentTimestamp + auctionDuration
  auctionInfo[tokenId].ended ← false
  NFSTInfo[tokenId].status ← "Idle"
  return True
bid // called by bidders
  input: tokenId, bid
  if(currentTimestamp > endTime or bid < highestBid or caller = highestBidder) then
      return False
  auctionInfo[tokenId]. highestBidder ← caller
  auctionInfo[tokenId].highestBid ← bid
  return True
end_auction // called by spectrum owner
  input: tokenId
  if(caller ≠ spectrum owner or currentTimestamp < endTime) then
      return False
  auctionInfo[tokenId].ended ← true
  for (i = 0; i < auctionInfo[tokenId].bidders.length; i++)
      bid ← auctionInfo[tokenId].pendingReturns[auctionInfo[tokenId].bidders[i]]
      auctionInfo[tokenId].pendingReturns[auctionInfo[tokenId].bidders[i]] = 0;
      auctionInfo[tokenId].bidders[i].send(bid)
  set_user(tokenId, highestBidder, lease duration)
  auctionInfo[tokenId].beneficiary.transfer(highestBid)
  return True
```

Fig. 6. NFST auction contract.

## B. Experimental Results Display

To enhance the user experience of the spectrum auction platform, we developed it using the Vue.js front-end framework [15]. We used the web3.js [16] library to facilitate user interaction between the spectrum auction platform and the auction smart contract. Fig. 7 shows the spectrum auction platform.

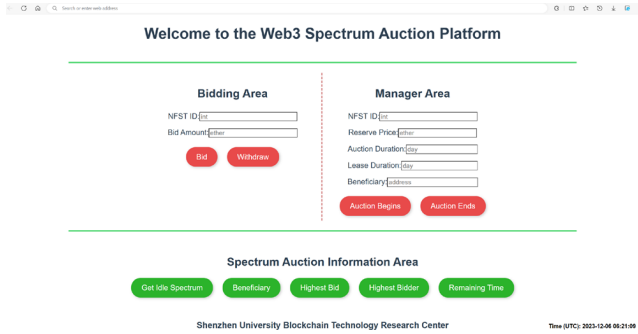

Fig. 7. Spectrum auction platform.

At the onset of the spectrum auction, the spectrum owner inputs key parameters: NFST ID, bidding duration, lease duration, beneficiary address, and starting price for the NFST being auctioned. The owner then initiates the auction process by calling the *startAuction* function with the "Auction Begins" button. Bidders can access the spectrum auction platform and submit bids using the NFST ID. As outlined in Section III, the procedure involves six selected bidders, including 3 SUs and 3 PUs, to participate in the spectrum auction. The experimental procedure can be summarized as:

a) Bidders access idle spectrum information from the blockchain by using the "Get Idle Spectrum" button, which provides details on available spectrum resources and their corresponding NFST IDs.

b) During bidding, participants enter the NFST ID to bid on the specific NFST. They can utilize the "Beneficiary" button to obtain the address of the idle spectrum beneficiary and verify its legitimacy. Using the "Highest Bid" button, participants can retrieve the current highest bid amount for the NFST to decide whether to continue participating in the auction. To place a bid, they input the bid amount into the "Bid Amount" field and then execute the bid function in the auction contract by clicking the "Bid" button to complete the bidding round. In this experiment, Ether serves as the circulating cryptocurrency, and the unit of measurement is in ether. Table 1 provides the account addresses and bid amounts for all six participants.

TABLE I. BIDDER ADDRESSES AND BID AMOUNTS

| bidders | addresses | amounts (ether) |
|---|---|---|
| $SU_1$ | 0x5B38Da6a70…5f56beddC4 | 2.0 |
| $SU_2$ | 0xAb8483F64d…3315835cb2 | 2.5 |
| $SU_3$ | 0x4B20993Bc4…A9e22C02db | 2.8 |
| $PuBuyer_1$ | 0x78731D3Ca6…8A495cabaB | 3.0 |
| $PuBuyer_2$ | 0x617F2E2fD7…c91465E7f2 | 3.1 |
| $PuBuyer_3$ | 0x17F6AD8Ef9…FE4348c372 | 3.5 |

c) At the end of the auction, the spectrum owner calls the *endAuction* contract via the "Auction ends" button to complete the auction for the spectrum lease term. Fig. 8 displays the final bidding results, with the winning bidder's account address indicated as 0x17F6AD8Ef982297579C203069C1DbfFE4348c372 and their bid amount listed as 3.5 ether (in addition to a small gas fee). The rest of the bidders can withdraw the bid amount paid for their previous participation in the auction by using the "Withdraw" button.

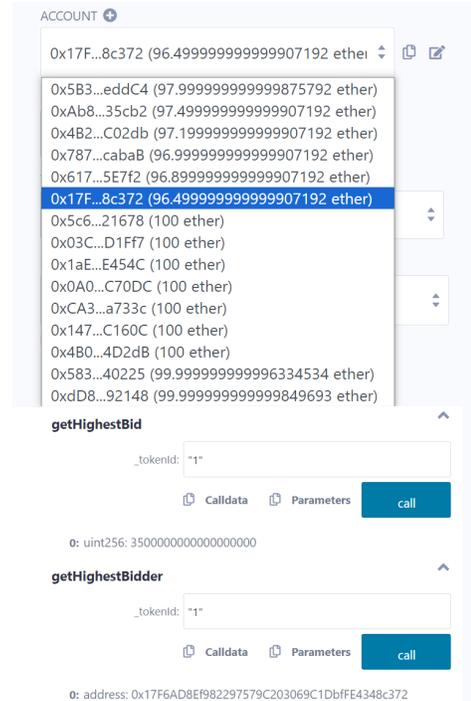

Fig. 8. Spectrum auction results.

d) Upon calling the *endAuction* contract, the spectrum owner calls the NFST user setting contract, as Section IV elaborates. This contract establishes the rental duration for the corresponding spectrum resource of the NFST. The winner's account address is set as the user address, and spectrum status is updated to "*Occupied*" during the lease duration, as depicted in Fig. 9.

```
[
  {
    "from": "0xfC713AAB72F97671bADcb14669248C4e922fe2Bb",
    "topic": "0x4e06b4e7000e659094299b3533b47b6aa8ad048e95e872d23d1f4ee55af89cfe",
    "event": "UpdateUser",
    "args": {
      "0": "1",
      "1": "0x17F6AD8Ef982297579C203069C1DbfFE4348c372",
      "2": "1703136913",
      "tokenId": "1",
      "user": "0x17F6AD8Ef982297579C203069C1DbfFE4348c372",
      "expires": "1703136913"
    }
  },
  {
    "from": "0xfC713AAB72F97671bADcb14669248C4e922fe2Bb",
    "topic": "0xdf5d830ac460106bcc950cc7c98ec13ba07ba7b34d1f143e6a9c5486e8fe9205",
    "event": "UpdateSpectrumStatus",
    "args": {
      "0": "1",
      "1": "Occupied",
      "tokenId": "1",
      "status": "Occupied"
    }
  }
]
```

Fig. 9. Set spectrum users and status.

## VI. Conclusion

In this paper, we proposed a blockchain-based DSS solution to address the high management costs and vulnerability of spectrum data in traditional centralized DSS systems. We introduced the ERC4907 protocol for minting rentable NFSTs as the unique identifiers for spectrum resources. Bidders can participate in auctions for leasing the NFSTs during specific periods, and the winner can utilize the corresponding spectrum resources within the lease period. The storage, retrieval, and transaction of spectrum data are seamlessly facilitated through blockchain and smart contracts, guaranteeing the reliability and security of the overall DSS system.

## Acknowledgment

China University Industry-Research-Innovation Fund[2022BL076], Shenzhen University Graduate Education Reform Research Project [SZUGS2023JG02], 2023 Shenzhen University Teaching Reform Research Project [JG2023097], Foundation of Shenzhen [20220809155455002], National Natural Science Foundation of China[62171291], Shenzhen Key Re-search Project[JSGG20220831095603007, JCYJ20220818100810023, JCYJ20220818101609021].

## References

[1] Ji Z, Liu K J R. Cognitive radios for dynamic spectrum access-dynamic spectrum sharing: A game theoretical overview[J]. IEEE Communications Magazine, 2007, 45(5): 88-94.

[2] Valenta V, Maršálek R, Baudoin G, et al. Survey on spectrum utilization in Europe: Measurements, analyses and observations[C]//2010 Proceedings of the fifth international conference on cognitive radio oriented wireless networks and communications. IEEE, 2010: 1-5.

[3] Mitola J, Maguire G Q. Cognitive radio: making software radios more personal[J]. IEEE personal communications, 1999, 6(4): 13-18.

[4] Kotobi K, Bilen S G. Secure blockchains for dynamic spectrum access: A decentralized database in moving cognitive radio networks enhances security and user access[J]. ieee vehicular technology magazine, 2018, 13(1): 32-39.

[5] Liang Y, Lu C, Zhao Y, et al. Interference-based consensus and transaction validation mechanisms for blockchain-based spectrum management[J]. IEEE Access, 2021, 9: 90757-90766.

[6] Zheng S, Han T, Jiang Y, et al. Smart contract-based spectrum sharing transactions for multi-operators wireless communication networks[J]. IEEE Access, 2020, 8: 88547-88557.

[7] Bhattarai S, Park J M J, Gao B, et al. An overview of dynamic spectrum sharing: Ongoing initiatives, challenges, and a roadmap for future research[J]. IEEE Transactions on Cognitive Communications and Networking, 2016, 2(2): 110-128.

[8] Wei WANG, Zuguang LI, Qihui WU. Dynamic spectrum sharing access technology based on blockchain[J]. Chinese Journal on Internet of Things, 2020, 4(2): 26-34.

[9] Ariyarathna T, Harankahadeniya P, Isthikar S, et al. Dynamic spectrum access via smart contracts on blockchain[C]//2019 IEEE Wireless Communications and Networking Conference (WCNC). IEEE, 2019: 1-6.

[10] "ERC-721: Non-Fungible Token Standard," Ethereum Improvement Proposals, no. 721, January 2018. [Online serial]. Available: https://eips.ethereum.org/EIPS/eip-721.

[11] "ERC-4907: Rental NFT, an Extension of EIP-721," Ethereum Improvement Proposals, no. 4907, March 2022. [Online serial]. Available: https://eips.ethereum.org/EIPS/eip-4907.

[12] Kumarathunga M, Neves Calheiros R, Ginige A. Sustainable microfinance outreach for farmers with blockchain cryptocurrency and smart contracts[J]. International Journal of Computer Theory and Engineering, 2022: 9-14.

[13] Remix IDE for Ethereum Smart Contract Programming, Sep. 2023, [Online] Available: https://remix.ethereum.org/.

[14] Buterin V. A next-generation smart contract and decentralized application platform[J]. white paper, 2014, 3(37): 2-1.

[15] Nelson B. Getting to know Vue.js[J]. Apress Media, August, 2018.

[16] Web3 Javascript API to Interact With Ethereum Nodes, Sep. 2023, [Online] Available: https://github.com/ethereum/wiki/wiki/JavaScript-API.